\documentclass[letterpaper, 10 pt, conference]{ieeeconf}
\usepackage[english]{babel}
\usepackage[utf8]{inputenc}
\usepackage{amsmath}
\usepackage{breqn}
\usepackage{filecontents,lipsum}
\usepackage[noadjust]{cite}
\usepackage{cite}
\usepackage{subcaption}
\usepackage{graphicx}
\usepackage[version=4]{mhchem}
\usepackage{siunitx}
\usepackage{longtable,tabularx}
\usepackage{caption}
\usepackage{hyperref}
\usepackage{float}
\usepackage{booktabs}
\usepackage{authblk}
\usepackage{amsfonts}
\usepackage{caption}
\usepackage{algorithm}
\usepackage{algpseudocode}
\usepackage{siunitx}
\usepackage{setspace}
\usepackage[export]{adjustbox}
\usepackage{multirow}
\usepackage{balance}
\usepackage{microtype}
\usepackage{graphics}

\def\p(#1|#2){p(#1\,|\,#2)}
\def\q(#1|#2){q(#1\,|\,#2)}

\newtheorem{rem}{Remark}

\renewcommand{\thetable}{\arabic{table}}
\algnewcommand{\Inputs}[1]{%
  \State \textbf{Inputs:} 
   \hspace*{0.3em}\parbox[t]{\linewidth}{\raggedright #1}
}
\algnewcommand{\Initialize}[1]{%
  \State \textbf{Initialize:}
   \hspace*{0.3em}\parbox[t]{.8\linewidth}{\raggedright #1}
}
\algnewcommand{\Output}[1]{%
  \State \textbf{Outputs:}
   \hspace*{0.3em}\parbox[t]{.8\linewidth}{\raggedright #1}
}
\title{\bf{Nonlinear Model Predictive Control Based on Constraint-Aware Particle Filtering/Smoothing}
}
\IEEEoverridecommandlockouts

\author[1]{Iman Askari\textsuperscript{1}, Shen Zeng\textsuperscript{2} and Huazhen Fang\textsuperscript{1}
\thanks{\textsuperscript{1}I. Askari and H. Fang are with the Department of Mechanical Engineering, University of
Kansas, Lawrence, KS 66045, USA (e-mail: {\tt\small askari, fang@ku.edu}).}
\thanks{\textsuperscript{2}S. Zeng is with the Department of Electrical and System Engineering, Washington University, St. Louis, MO 63130, USA (e-mail: {\tt\small s.zeng@wustl.edu}) and was supported in part by the NSF grant CMMI-1933976.}

}
\begin{document}

\maketitle
\begin{abstract} Nonlinear model predictive control (NMPC) has gained widespread use in many applications. Its formulation traditionally  involves repetitively solving a nonlinear constrained optimization problem online. In this paper, we investigate NMPC   through the lens of Bayesian estimation and highlight that the Monte Carlo sampling method can offer a favorable way to implement NMPC.  We develop  a constraint-aware particle filtering/smoothing method and exploit it to  implement NMPC.  The new sampling-based NMPC algorithm  can be executed easily and efficiently even for complex nonlinear systems, while potentially mitigating  the issues of computational complexity and local minima faced by numerical optimization in conventional studies.   The effectiveness of the proposed algorithm is evaluated through a simulation study.
\end{abstract}
 
\section{Introduction}
Nonlinear model predictive control (NMPC) has emerged as one of the most important control methods, with numerous applications ranging from process control and robotics to energy management and traffic control~\cite{Allgower:Birkhauser:2000, Qin:CEP:2003, Rakovic:Birkhauser:2019}. Its immense success is owed to its ability of predictively optimizing the control of a nonlinear system subject to constraints. At the center of its formulation, NMPC is designed to address an optimal control problem repetitively at every time instant in a receding-horizon manner~\cite{Johansen:NTNU:2011}. Its implementation hence hinges on the online solution of a nonlinear constrained optimization problem. A simple approach is to do model linearization and solve the resultant linear model predictive control problem by a Newton-type method, if the system is subject to only linear constraints~\cite{Li:IECR:1988, Brengel:IECR:1989}. Further, nonlinear programming must be considered when generic nonlinear constraints are present. Two popular classes of methods to deal with it include the sequential quadratic programming and nonlinear interior-point methods~\cite{Diehl:Springer:2009,Rawlings:NHP:2017}. As practical applications often demand computational efficiency, significant research efforts have recently been dedicated to NMPC driven by fast real-time optimization. The continuation/GMRES method and various other proposed approaches of leveraging the structures of NMPC have proven useful in improving the computational speed~\cite{Ohtsuka:AUTO:2004,Shen:TCST:2017,Wang:TCST:2010,Biegler:Birkhauser:2000,Rawlings:NHP:2017}. Evolutionary algorithms, e.g., the particle swarm optimization method, have also received some attention in the literature~\cite{pso,Du:TCST:2000}. This is mainly due to their ability to achieve global optimization for nonconvex nonlinear programs, even though they are computationally expensive.

Despite the advancements, the high computational costs of nonlinear constrained optimization remains a bottleneck for the application of NMPC, especially when it comes to high-dimensional or highly nonlinear systems. Another concern lies in whether the global minimum can be attained, given that the optimization is nonconvex in many cases. The particle filtering approach, as a sequential Monte Carlo sampling method, has shown effective in handling the complexity of nonlinear systems in state estimation tasks~\cite{seqMC,sarkka}. Its utility can be extended to tackle the above issues facing NMPC. The study in~\cite{STAHL} illustrates particle filtering as a means to implement NMPC, by estimating the optimal control input sequence from the reference signal to track over the receding horizon. However, it is the only work on this subject to our knowledge, though a few other studies use particle filtering for the purpose of state estimation in output-feedback NMPC~\cite{Botchu:IFAC:2007,Blackmore:AIAA:2006}. Another related line of research is to solve NMPC using the Monte Carlo sampling-based optimization in place of numerical optimization~\cite{Kantas2009, MCMC}.

It is noteworthy that the method offered in~\cite{STAHL} adopts an existing particle filter algorithm and thus cannot deal with generic state-input inequality constraints despite their presence in various practical systems. In this work, we aim to overcome this limitation and make the following contributions:
\begin{itemize}

\item We introduce a systematized formulation of NMPC through the lens of Bayesian estimation  and an     implementation based on vanilla particle filtering/smoothing. 

\item Further, we propose a novel  constraint-aware particle filtering/smoothing approach using the barrier function method and then apply it to develop a new NMPC algorithm.
 
\end{itemize}
The proposed algorithm makes a crucial advancement of the particle-filtering-based NMPC  and can find   prospective use in a broad range of applications. 

The rest of the paper is organized as follows. Section~\ref{NMPC-Bayesian-Lens} interprets the NMPC problem as a Bayesian estimation problem. Section~\ref{NMPC-PF} presents a realization of NMPC based on particle filtering/smoothing and proposes the constraint-aware particle filtering/smoothing method to implement NMPC. Section~\ref{num-sim} evaluates the proposed algorithm through a simulation example. Finally, Section~\ref{conc} concludes the paper.

\section{NMPC through the Lens of Bayesian Estimation} \label{NMPC-Bayesian-Lens}
In this section, we present the NMPC problem and then examine it from the perspective of Bayesian estimation.

 Consider the following discrete-time nonlinear system: 
\begin{align}\label{state-equation}
x_{k+1} = f(x_{k},u_{k}),
\end{align}
where $x_{k} \in \mathbb{R}^{n_x}$ is the system state, $u_{k} \in \mathbb{R}^{n_u}$ is the control input, and the mapping $f:  \mathbb{R}^{n_x} \times \mathbb{R}^{n_u} \rightarrow \mathbb{R}^{n_x}$ represents the state transition function.  The system is subject to the following inequality constraints:
\begin{align}\label{inequality-constraints}
g_j(x,u)\leq 0,  \ \forall j=1,\ldots,m,
\end{align}
where $m$ is the total number of constraints. We suppose that   the control objective is to make $x_k$ track a reference signal $r_k$. A corresponding  NMPC problem can be stated as follows:
\begin{subequations} \label{NMPC-Standard}
\begin{align}
\min_{u_{k:k+H}} \quad &  \sum_{t=k}^{k+H}
 \left\| r_t -  x_t\right\|_Q^2 +\left\| u_t \right\|_R^2  ,\\
\mathrm{s.t.} \quad & x_{t+1} = f(x_{t},u_{t}) , \\
& g_j(x_t,u_t) \leq 0 \quad \forall j=1,\ldots,m,\\ \nonumber
& t = k, \ldots, k+H,
\end{align}
\end{subequations}
where $H$ is the length of the upcoming horizon,  $u_{k:k+H} = \left\{u_k, u_{k+1}, \ldots, u_{k+H}\right\}$,   $Q$ and $R$  are weighting matrices, and $x_k$ is known. The above problem seeks  to determine the optimal control input sequence $u_{k:k+H}^\ast$ to minimize an objective function, which is a weighted quadratic sum of the    control cost and tracking error over $H$ steps into the future. The literature has conventionally resorted to numerical optimization approaches to compute $u_{k:k+H}^\ast$. Once the optimization is done, the first element  of   $u_{k:k+H}^\ast$, $u_k^\ast$, will be applied to control the system, with the rest  discarded. The same   optimization and control procedure will repeat itself recursively at the future time instants.  
To sum up,     NMPC, at its core, pursues model-based predictive optimization of a system's operation in a receding horizon. 

Another perspective to investigate     NMPC   is based on optimal estimation.  The overarching  notion is  to interpret the NMPC problem as a problem of estimating $x_{k:k+H}$ and $u_{k:k+H}$ from  $r_{k:k+H}$. Specifically, we can view  $x_{k:k+H}$ and $u_{k:k+H}$ together as the state characteristic of a  virtual dynamic system and   $r_{k:k+H}$ as the   virtual measurements made on this system. The virtual system can   be expressed as 
\begin{align}\label{Virtual-System}
\left\{
\begin{aligned}
x_{t+1} &= f(x_t, u_t),\\
u_{t+1} &=  w_t,\\
r_t &= x_t+ v_t,
\end{aligned}
\right. 
\end{align}
for $t=k, \ldots, k+H$, where $w_t$ and $v_t$ are additive disturbances. Given~\eqref{Virtual-System}, we can consider a moving horizon estimation (MHE) problem~\cite{Rao:TAC:2003} to estimate the combined state $x_t$ and $u_t$:
\begin{subequations} \label{MHE}
\begin{align}
\min_{x_{k:k+H}, u_{k:k+H}} \quad &  \sum_{t=k}^{k+H}
  \left\| v_t \right\|_Q^2 + \left\| w_t \right\|_R^2,\\
\mathrm{s.t.} \quad & x_{t+1} = f(x_{t},u_{t}) , \\
& u_{t+1} = w_t,  \\
& r_t = x_t+ v_t,
\end{align}
\begin{align}
& g_j(x_t,u_t) \leq 0 \quad \forall j=1,\ldots,m,\\ 
& t = k, \ldots, k+H. \nonumber
 \end{align}
\end{subequations} 
The formulation in~\eqref{MHE} is equivalent to~\eqref{NMPC-Standard}, suggesting the viability of treating NMPC as  an estimation problem. This view opens up a different way of dealing with NMPC.

Bayesian estimation offers another   means of performing state estimation for~\eqref{Virtual-System}. 
In principle, we consider the probability distribution    $\p(x_{k:k+H}, u_{k:k+H} | r_{k:k+H})$, which characterizes the  information that  $r_{k:k+H}$ contains about  $x_{k:k+H}$ and $u_{k:k+H}$. Using   Bayes' rule, we have
\begin{align}\label{Bayesian-state-input-estimation}\nonumber
&  \p(x_{k:k+H}, u_{k:k+H} | r_{k:k+H}) \\ \nonumber
&    \propto  \prod_{t=k}^{k+H}  \p(r_{t } | x_{t })    \p(u_t)   \cdot    \prod_{t=k}^{k+H-1} \p(x_{t+1} | x_t, u_t)  \cdot   p(x_k)    \\
&  =\prod_{t=k}^{k+H}  \p(r_{t } | x_{t })       p(u_t)   ,
\end{align}
where the last equality holds because  $x_k$ is known and  $x_{t+1} = f(x_t, u_t)$ is deterministic. By~\eqref{Bayesian-state-input-estimation}, if  assuming $ \p(r_{t } | x_{t })  \sim \mathcal{N} \left(x_t, Q^{-1}\right)$ and $p(u_t) \sim \mathcal{N} \left(0, R^{-1} \right)$ and considering the maximum likelihood estimation based on 
\begin{align*}
\max_{x_{k:k+H}, u_{k:k+H}} \log \p(x_{k:k+H}, u_{k:k+H} | r_{k:k+H}),
\end{align*}
we will obtain the same formulation as in~\eqref{MHE} except without the inequality constraints.   

Bayesian estimation encompasses different ways of implementation. Among them, particle filtering is one of the most powerful approaches, which uses sequential Monte Carlo sampling to approximate the conditional probability distribution of a system's  state given the measurements. The study in~\cite{STAHL} leverages particle filtering  to develop an NMPC method. However, by design, it is unable to handle the    state-input inequality constraints~\eqref{inequality-constraints}, thus limiting its applicability to many practical systems.   To overcome this limitation, we will propose a constraint-aware particle filtering method and exploit to enable NMPC, with details to be offered in the next sections. 

\begin{rem}
The execution of NMPC has conventionally relied on  numerical optimization. However, for  high-dimensional or highly nonlinear systems,   the   computational complexity  can be extremely high and even prohibitive, and the implementation (e.g., coding and testing) rather burdensome. By contrast, Monte Carlo sampling is computationally more efficient and easier to implement in such scenarios. Some recent studies have also pointed out that sampling can be a more favorable choice than numerical optimization for   large-scale tasks~\cite{Ma-Chen-Flammarion-Jordan:PNAS:2019}. Sampling-based NMPC hence holds significant promise for controlling complex systems that emerge in various problem domains. 
\end{rem}

\section{NMPC via Constraint-Aware Particle Filtering/Smoothing}\label{NMPC-PF}

In this section, we first present the problem of addressing NMPC via particle filtering/smoothing and  show a basic approach to this end. Then, we develop the constraint-aware particle filtering/smoothing and apply it to NMPC with generic inequality constraints. 

\subsection{Problem Formulation}

We consider the NMPC-oriented virtual system in~\eqref{Virtual-System} and rewrite it compactly as
\begin{align}\label{Virtual-System-Compact}
\left\{
\begin{aligned}
\bar{x}_{t+1} &= \bar{f}(\bar{x}_t) + \bar{w}_t, \\ 
r_t &=M \bar{x}_t  +v_t,
\end{aligned}
\right.
\end{align}
for $t = k, \ldots, k+H$, 
where $\bar{x}_t=\left[x_{t}^{\top} \ u_{t}^{\top}\right]^{\top}$, $\bar{w}_t=\left[0^\top \ w_t^{\top}\right]^{\top}$,   $\bar f$ stems from $f$, and  $M = \left[ I \ 0\right]$. For now, we neglect the inequality constraints and will come back to this point in Section~\ref{NMPC-Constrained-PF/PS}. As discussed in Section~\ref{NMPC-Bayesian-Lens}, the original NMPC problem could be cast as estimating $\bar x_{k:k+H} $ from $r_{k:k+H}$.  According to the principle of Bayesian estimation, it is of interest to determine $\p(\bar x_{k:k+H} | r_{k:k+H})$.  We hence consider the following   recurrence relation
\begin{multline}\label{Bayesian-filtering}
\p(\bar{x}_{k:t}|r_{k:t}) \propto \p(r_{t} |\bar{x}_{k:t} ) \p(\bar{x}_{t} |\bar{x}_{t-1}) 
\p(\bar{x}_{k:t-1}|r_{k:t-1}), 
\end{multline}

for $t = k, \ldots, k+H$. This relation shows  a forward pass from $\p(\bar{x}_{k:t-1} | r_{k:t-1})$ to $\p(\bar{x}_{k:t} | r_{k:t})$ and is known as {\em filtering}. 
Since NMPC only applies the first element of the computed control input sequence, we further need to consider $\p(\bar{x}_k | r_{k:k+H})$.   A  backward   recursion as follows is necessary to this end:
\begin{multline} \label{Bayesian-smoothing}
p(\bar{x}_t|r_{k:k+H}) = p(\bar{x}_t|r_{k:t}) \int \frac{p(\bar{x}_{t+1}|\bar{x}_t)}{p(\bar{x}_{t+1}|r_{k:t})} \\  \times   p(\bar{x}_{t+1}|r_{k:k+H})d\bar{x}_{t+1},  
\end{multline}
for $t=k+H, \ldots, k$,  which describes a backward pass  from $\p(\bar{x}_{t+1} | r_{k:k+H})$ to $\p(\bar{x}_{t} | r_{k:k+H})$ and is known as {\em smoothing}. 
 Together, \eqref{Bayesian-filtering}-\eqref{Bayesian-smoothing}  form a Bayesian forward filtering/backward smoothing framework to realize NMPC. However, a closed-form solution to $p(\bar{x}_k|r_{k:k+H}) $ is generally intractable to derive  for nonlinear systems. This motivates the use of sequential Monte Carlo sampling, which  leads to particle filtering/smoothing.

\subsection{NMPC via Vanilla  Particle Filtering/Smoothing}\label{NMPC-Vanilla-PF/S}

To begin with, we discuss the forward filtering and consider $\p (\bar x_{k:t} | r_{k:t})$ for $k \leq t \leq k+H$. Given that an analytical expression is unavailable for $\p (\bar x_{k:t} | r_{k:t})$, we seek to approximate it by developing a sample-based empirical distribution. Nonetheless, it is non-trivial to draw the samples since $\p (\bar x_{k:t} | r_{k:t})$ is unknown. To deal with this problem, one can leverage the technique of importance sampling. The main idea lies in drawing samples from an alternative known distribution $\q (\bar x_{k:t} | r_{k:t})$, which is called   importance distribution, and then evaluating the weights of the samples in relation to $\p (\bar x_{k:t} | r_{k:t})$. Suppose that $N$ samples, $\bar x_{k:t}^i$ for $i=1,\ldots,N$, are drawn from $\q (\bar x_{k:t} | r_{k:t})$ . Their importance weights are then given by
\begin{align}\label{Weights}
W_t^i = \frac{\p (\bar x_{k:t} | r_{k:t})}{\q (\bar x_{k:t} | r_{k:t})}.
\end{align}
If $W_t^i$ for $i=1,\ldots,N$ are normalized to be between 0 and 1, then $\p (\bar x_{k:t} | r_{k:t})$ is approximated as
\begin{align*}
\p (\bar x_{k:t} | r_{k:t}) \approx \sum_{i=1}^N W_k^i \delta \left( \bar x_{k:t} - \bar x_{k:t}^i\right). 
\end{align*}
Note that~\eqref{Weights} can be written as
\begin{align*}
W_t^i = \frac{\p (\bar x_{k:t} | r_{k:t})}{\q (\bar x_{k:t} | r_{k:t})} = \frac{\p (  r_{t} | \bar x_t^i) \p(\bar x_t^i | \bar x_{t-1}^i)}{\q (\bar{x}_t^i | \bar{x}_{t-1}^i, r_{k:t} )} W_{t-1}^i,
\end{align*}
which suggests a recursive update of $W_t^i $.  

There are different ways to select the importance distribution $q$. In general, it should be chosen such that it is well related with the target distribution $p$ and allows samples to be easily drawn. A straightforward choice is to let $\q (\bar{x}_t | \bar{x}_{t-1}, r_{k:t} ) =  \p (\bar{x}_t | \bar{x}_{t-1})$. This implies that, at time $t$, we can draw samples  $\bar x_t^i \sim \p(\bar x_t | \bar x_{t-1}^i)$, with associated normalized weights computed by
\begin{align} \label{tracking-weight}
W_t^i = \frac{\p(r_t | \bar x_t^i)}{\sum_{j=1}^N \p(r_t | \bar x_t^j)}.
\end{align}
The resulting implementation is called  the {\em bootstrap particle filter}. A common challenge in particle filtering   is   particle degeneracy, where a majority of the particles have zero or almost zero weights after a few time steps. The overall quality of the particles become extremely low and reduce the estimation performance. A useful method to resolve this problem is resampling, which  replaces low-weight particles with those that have high weights \cite{resample}. 

Then, we perform the backward smoothing, which will provide a more accurate estimation of $\bar x_k$. Rewrite~\eqref{Bayesian-smoothing} as:
\begin{multline} \label{Bayesian-smoothing-new}
\p(\bar{x}_t|r_{k:k+H}) = \p(\bar{x}_t|r_{k:t})      \int \frac{\p(\bar{x}_{t+1}|\bar{x}_t)}{\int   \p(\bar{x}_{t+1}|\bar x_{t}) \p(\bar{x}_{t}|r_{k:t})   d \bar x_t} 
\\  \times \p(\bar{x}_{t+1}|r_{k:k+H})d\bar{x}_{t+1},  
\end{multline}
where the relation $\p (\bar x_{t+1} | r_{k:t}) = \int   \p(\bar{x}_{t+1}|\bar x_{t}) \p(\bar{x}_{t}|r_{k:t})   d \bar x_t$ is used. It is noted that all the probability distributions in~\eqref{Bayesian-smoothing-new} can be approximated by  empirical distributions based on samples computed in the filtering procedure. Hence, according to~\eqref{Bayesian-smoothing-new}, we only need to reweight the samples in smoothing:
\begin{align}\label{reweighting}
W_{t|k+H}^i = \sum_{j=1}^N W_{t+1|k+H}^i\frac{W_t^i \ p(\bar{x}_{t+1}^j|\bar{x}_t^i)}{ \sum_{l=1}^N W_t^l \p(\bar{x}_{t+1}^j|\bar{x}_t^l)},
\end{align}
where   $W_{k+H|k+H}^i = W_{k+H}^i$. The resultant procedure is called  {\em reweighted particle smoother}. The smoothing will end up with   the following empirical distribution for $p(\bar{x}_k|r_{k:k+H})$:
\begin{equation*}
\p(\bar{x}_k|r_{k:k+H})\approx\sum_{i=1}^N W_{k|k+H}^i\delta(\bar{x}_k-\bar{x}_k^i).
\end{equation*}
This implies that the best estimate of $\bar x_k$ from $r_{k:k+H}$ is
\begin{align}\label{estimate}
\hat{\bar x}_k^\ast = \sum_{i=1}^N W_{k|k+H}^i \bar x_k^i,
\end{align}
which then gives the optimal control input at time $k$.

The above introduces a vanilla particle filtering/smoothing algorithm  to implement NMPC. It bears a resemblance to the algorithm in~\cite{STAHL} but uses the  reweighted  particle smoother to enhance the accuracy of smoothing. 
However,  both algorithms, at their core, do not take the inequality constraints into account. Next, we  propose a constraint-aware particle filtering/smoothing approach to remedy this issue. 

\begin{figure}[t]
	  \centering
	  \includegraphics[width=.8\linewidth, trim={0.3cm 0cm 0.3cm 0.3cm},clip]{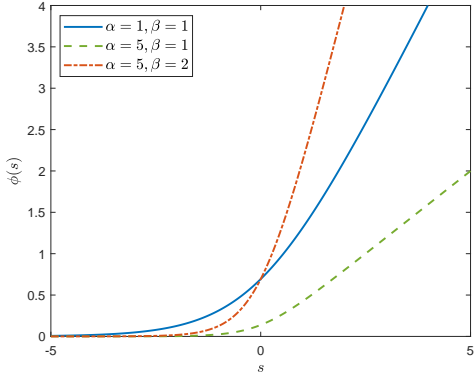}  
	  \caption{The softplus barrier function for different values of $(\alpha, \beta)$.}
	  \label{fig:barrier}
\end{figure}
\subsection{Constraint-Aware Particle Filtering/Smoothing for NMPC} \label{NMPC-Constrained-PF/PS}

Here, we develop an effective method to enable particle filtering/smoothing with an awareness of the inequality constraints. The idea lies in leveraging a barrier function to create  virtual measurements about the constraint satisfaction and then incorporate them into the estimation procedure. 

We first construct a virtual measurement equation as
\begin{align}\label{Virtual-measurement-barrier-function}
z_t = \phi\left(g(\bar{x}_t)\right) + \eta_t,
\end{align}
where $z$ is the virtual measurement variable quantifying  the constraint satisfaction, $g$ is the collection of $g_j$ for $j=1,\ldots, m$, $\eta$ is an additive small noise, and $\phi$ is a barrier function. Then,~\eqref{Virtual-measurement-barrier-function} will be used to evaluate the   weights of the particles and penalize those failing to satisfy the constraints. 

In general, a barrier function outputs   almost zero at a point within a constraint set,  almost infinity  near the inner boundary of the set, and infinity outside the set~\cite{borrelli_bemporad_morari_2017}. However, such a barrier function will  assign just zero weights to samples that violate the constraints. This may further deteriorate  the particle degeneracy issue, impoverishing the entire ensemble of samples.  Meanwhile, from the viewpoint of estimation,  one only needs to make the final aggregated state estimate meet the constraints. 
Therefore, we   depart from the traditional way to  consider a fully continuous barrier function, which is  designed to output almost zero within the constraint set and very large values outside the set.  To this end, we choose  the softplus function as a barrier function, which is expressed as
\begin{equation}
 \phi{\left( s \right)} = \frac{1}{\alpha}\ln\left(1+\exp(\beta \cdot s)\right),
\end{equation}
where  $\alpha$  and $\beta$ are parameters to adjust the effect of the constraint violation on the particle weights. Fig.~\ref{fig:barrier} shows the shape of the function under different choices of $\alpha$ and $\beta$. The virtual measurement $z_t$ is accordingly set to be $0$. 

Now, let us add~\eqref{Virtual-measurement-barrier-function} to~\eqref{Virtual-System-Compact} and consider $\p( \bar x_{k:t} | r_{k:t}, z_{k:t})$ for constraint-aware filtering. Assuming that $\p(r_t, z_t | \bar x_t) = \p(r_t| \bar x_t) \p(z_t | \bar x_t)$, we can remold~\eqref{Bayesian-filtering} as
\begin{multline*}
\p(\bar{x}_{k:t}|r_{k:t},z_{k:t}) \propto \p(r_t|\bar{x}_t) \p(z_t|\bar{x}_t)    \p(\bar{x}_t|\bar{x}_{t-1}) \\ \cdot
\p(\bar{x}_{k:t-1}| r_{k:t-1},z_{k:t-1}).
\end{multline*}
Then, the bootstrap particle filter described in Section~\ref{NMPC-Vanilla-PF/S} can then be modified. Specifically, we still draw samples $\bar x_t^i \sim \p(\bar x_t | \bar x_{t-1}^i)$ for $i=1,\ldots,N$ at time $t$ but compute the weights via
\begin{align}\label{const_weight}
W_t^i = \frac{\p(r_t | \bar x_t^i) \p(z_t | \bar x_t^i)}{\sum_{j=1}^N \p(r_t | \bar x_t^i)  \p(z_t | \bar x_t^j)}.
\end{align}
With this change, the bootstrap particle filter diminishes the weights of the samples in violation of the constraints. Once the constraint awareness has been infused into the filtering procedure, we can apply the reweighted particle smoother without change. This can be seen from the fact that~\eqref{reweighting} will not change if we consider $\p(\bar x_t | r_{k:k+H}, z_{k:k+H})$ for the backward smoothing. Summarizing the above, we can readily formulate the new sampling-based NMPC algorithm based on the proposed constraint-aware particle filtering/smoothing method,   which is named as {CAP-NMPC} outlined in Algorithm~\ref{NMPC-CA-PF/S}.

\begin{rem}
Particle filtering for constrained   systems  has received attention in several studies. However, the proposed barrier function method offers two distinct benefits. First, 
using  the suggested barrier function, the particle filtering does not simply discard the particles violating the constraints and instead include the constraint satisfaction or violation into the weighting process.  This will avoid complicating  the  particle degeneracy problem and   balance with the need of approximating the state's posterior distribution. The treatment contrasts with the acceptance/rejection method in~\cite{lang}. Second, the computational cost will see only very mild increase, even when the number of the constraints is large  or the constraints are highly nonlinear,  as opposed to the constrained optimization method in~\cite{copt}. Finally, we highlight that the   barrier function method can be easily integrated into almost every particle filter algorithm for broader application.
\end{rem}

\begin{algorithm}[t]
\fontsize{10}{10}
  \caption{CAP-NMPC: NMPC Based on Constraint-Aware Particle Filtering/Smoothing}	  \label{NMPC-CA-PF/S}
  \begin{algorithmic}[1]
\State {Set up   NMPC}   by specifying the dynamic system~\eqref{state-equation}, the inequality constraints~\eqref{inequality-constraints}, the reference signal $r_{1:T}$, and the weighting matrices $Q$ and $R$

\State {Recast NMPC} as particle filtering/smoothing by setting up the virtual system~\eqref{Virtual-System-Compact} and~\eqref{Virtual-measurement-barrier-function}, and specifying $p(\bar x_k)$, $\p(\bar x_{k+1} | \bar x_k)$, $\p(r_k | \bar x_k)$ and $\p(z_k | \bar x_k)$

\For{$k=1, \ldots, T$}

\item[]{\hspace{14pt}\bf\em    Forward filtering}
 
\For{$t = k,\ldots, k+H$}

\If{$t=k$}

\State Draw samples $\bar x_k^i \sim  p(\bar x_k)$, $i=1, \ldots, N$

\Else

\State Draw samples $\bar x_t^i \sim \p(\bar x_t | \bar x_{t-1}^i)$, $i=1,\ldots,N$

\State Evaluate sample weights via~\eqref{const_weight}

\State Do resampling based on the weights

\EndIf

\EndFor

\item[]{\hspace{14pt}\bf\em    Backward smoothing}

\For{$t = k+H,\ldots, k$}

\If{$t=k+H$}

\State Assign $W_{k+H|k+H}^i = W_{k+H}^i$, $i=1,\ldots,N$

\Else 

\State Reweight the particles using~\eqref{reweighting}
\EndIf

\EndFor

\State Compute the optimal estimation of $\hat{\bar {x}}_k^\ast$ via~\eqref{estimate}

\State Export $u_k^*$, and apply it to the system~\eqref{state-equation}

\EndFor

\end{algorithmic}
\end{algorithm}

\section{Numerical Simulation}\label{num-sim}

In this section, we study the performance of the proposed algorithm for a path following problem of an autonomous vehicle. The code is available at: \url{https://github.com/KU-ISSL/PF_NMPC_ACC21}. 
The considered vehicle motion model is the nonlinear bicycle kinematic model, described in \cite[p.~27]{rajmani}:
\begin{subequations}
 \begin{align}
x^{p}_{k+1} &= x^{p}_{k} + \Delta{t}\cdot \nu_{k}\cos(\psi_{k}+\beta_{k}), \\
y^{p}_{k+1} &= y^{p}_{k} + \Delta{t}\cdot \nu_{k}\sin(\psi_{k}+\beta_{k}),  \\
\nu_{k+1} &= \nu_{k} + \Delta{t}\cdot a_{k},  \\
\psi_k &= \psi_{k-1} + \Delta{t}\cdot \frac{\nu_{k}}{l_r}\sin(\beta_{k}), 
\end{align}
\begin{align}
\beta_k &=\tan^{-1}\left(\frac{l_r}{l_r+l_f}\tan(\delta_{f,k})\right),
\end{align}
\end{subequations}
where $x^p_k$ and $y^p_k$ denote the position of the center of mass, $\nu_k$ is the vehicle's speed, $\psi_k$ is its heading angle, and $\beta_k$ is the side-slip angle. The lengths $l_r$ and $l_f$ represent the distance from the rear and front axles to the center of mass. The control sequence used to control the vehicle is ${u}_k=\left[a_k \ \delta_{f,k}\right]^{\top}$, where $a_k$ and $\delta_{f,k}$ are the acceleration and front-wheel steering angle, respectively. For the purpose of comparison, we run the vanilla particle filtering/smoothing approach in Section~\ref{NMPC-Vanilla-PF/S} as a benchmark, which is labeled as P-NMPC for convenience. To ensure the fairness of comparison, we require that both CAP-NMPC and P-NMPC use the same set of particles with $N = 100 $ throughout the simulation run. The setting used for the simulation is as follows:
\begin{figure}[t!]
\begin{subfigure}{.5\textwidth}  
  \centering
  \includegraphics[width=.76\linewidth, trim={0.5cm 0cm 0.5cm 0.5cm},clip]{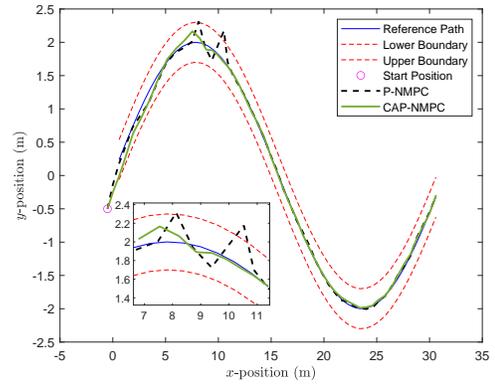}  
  \caption{Trajectories of the autonomous vehicle.}
  \label{fig:traj}
\end{subfigure}
\begin{subfigure}{.5\textwidth} 
  \centering
  \includegraphics[width=.76\linewidth, trim={0.5cm 0cm 0.5cm 0.5cm},clip]{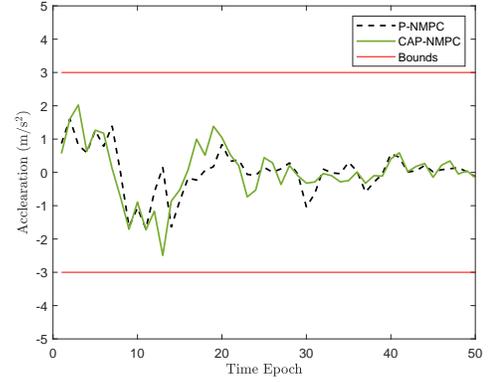}  
  \caption{Acceleration controls applied to the autonomous vehicle.}
  \label{fig:acc}
\end{subfigure}
\begin{subfigure}{.5\textwidth}\label{steer}
  \centering
  \includegraphics[width=.76\linewidth, trim={0.5cm 0cm 0.5cm 0.5cm},clip]{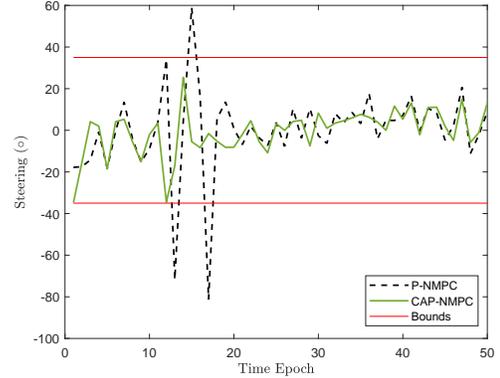}  
  \caption{Steering controls applied to the autonomous vehicle.}
  \label{fig:steer}
\end{subfigure}
\caption{Autonomous vehicle path following results.}
\label{RESULT}
\end{figure}

\begin{itemize}
\item The vehicle has to follow a sinusoidal track with the $x$-position ($x^{\text{track}}$) ranging from $0$ to $33 \: \text{m}$ with steps size $dx^{\text{track}} = 0.6 \: \text{m}$, and the $y$-position as $y^{\text{track}} =2\sin(0.2x^{\text{track}})$;
\item The vehicle is initialized from the state $x_0 = [-0.5, \ -0.5, \ 3, \ \pi/4]^{\top}$, where a deviation from the desired trajectory is introduced to better evaluate the performance of the reference tracking control;
\item The control input is constrained between the lower and upper bounds, $\underline{u} = [-3\ \mathrm{m/s^2},\ -35^{\circ}]^{\top}$ and $\overline{u} = [3\ \mathrm{m/s^2}, \ 35^{\circ}]$, respectively;
\item The state is constrained by the track boundaries, where the lower and upper track boundaries are each $0.3\: \mathrm{m}$ apart from the center of the track;
\item The disturbance $\bar{w}_k$, $v_k$, and $\eta_k$ are assumed to have Gaussian distributions with covariances as $Q_{\bar{w}} =\mathrm{diag}(0,0,0,0,0.8,0.4)$,  $Q_{v}= \mathrm{diag}(0.01,0.01,0,0,0,0)$, and $Q_{\eta}=0.01\textit{I}_5$, respectively. This setting corresponds to having $R=\mathrm{diag}(1.25,2.5)$ and $Q=\mathrm{diag}(100,100)$ for the NMPC problem in~\eqref{MHE};
\item The matrices $Q_{\bar{w}}$ and $Q_v$ are slightly perturbed to ensure positive definiteness for numerical stability;
\item The horizon length $H = 4$;
\item The barrier function parameters are set as $\alpha= 5$ and $\beta = 3$;
\item The axles are equidistant from the center of mass, with $l_r=l_f=0.5\: \mathrm{m}$.
\end{itemize}
\begin{table} 
\centering
\caption{RMSE and control cost achieved.}
        \begin{tabular}{@{}lll@{}}
        \toprule
         {}& RMSE & Cost \\
\midrule
          P-NMPC &0.330&1947\\
          \midrule
          CAP-NMPC &0.324&1862\\
        \bottomrule
        \end{tabular}
        \label{tab:result-table}
\end{table}

The root mean squared error (RMSE) is used to evaluate the tracking performance of the vehicle. 
The vehicle's trajectory is depicted in Fig.~\ref{fig:traj} with the zoomed region in the sub-axes showing an instance where the CAP-NMPC is able to satisfy the constraints while P-NMPC fails. Bearing in mind that both algorithms use the same set of particles for every horizon, it becomes evident from Fig~\ref{RESULT} that the CAP-NMPC has successfully managed to finely tune the particle weights so that the vehicle can run while staying within the pre-set trajectory boundaries. Similarly, the control acceleration and steering inputs have also satisfied the constraints in the CAP-NMPC case, as shown in Figs.~\ref{fig:acc}-\ref{fig:steer}. The control performance is evaluated in Table~\ref{tab:result-table}. The CAP-NMPC outperforms the P-NMPC for both metrics. This is because it maintains an awareness of constraints. To summarize, the results show that CAP-NMPC outperforms the P-NMPC, indicating that the CAP-NMPC can sufficiently and successfully implement NMPC.
\section{Conclusion}\label{conc}
In this paper, we have examined NMPC from the viewpoint of Bayesian estimation and leveraged particle filtering to estimate the optimal control input of an NMPC problem. Compared with the commonly used numerical optimization, this treatment exploits Monte Carlo sampling to improve the efficiency, effectiveness, and easiness of implementing NMPC, especially conducive to the control of nonlinear systems. To endow the particle-filtering-based NMPC with the capability of handling generic state-input inequality constraints, we developed a constraint-aware particle filtering/smoothing approach based on the barrier function method. We evaluated the proposed NMPC algorithm via a simulation study of path following for autonomous vehicles. Our future work will include applying the algorithm to more complex nonlinear systems and using more sophisticated particle filtering to enhance the algorithm.

\balance
\bibliographystyle{IEEEtran}
\bibliography{cit}

\begin{thebibliography}{10}
\providecommand{\url}[1]{#1}
\csname url@samestyle\endcsname
\providecommand{\newblock}{\relax}
\providecommand{\bibinfo}[2]{#2}
\providecommand{\BIBentrySTDinterwordspacing}{\spaceskip=0pt\relax}
\providecommand{\BIBentryALTinterwordstretchfactor}{4}
\providecommand{\BIBentryALTinterwordspacing}{\spaceskip=\fontdimen2\font plus
\BIBentryALTinterwordstretchfactor\fontdimen3\font minus
  \fontdimen4\font\relax}
\providecommand{\BIBforeignlanguage}[2]{{%
\expandafter\ifx\csname l@#1\endcsname\relax
\typeout{** WARNING: IEEEtran.bst: No hyphenation pattern has been}%
\typeout{** loaded for the language `#1'. Using the pattern for}%
\typeout{** the default language instead.}%
\else
\language=\csname l@#1\endcsname
\fi
#2}}
\providecommand{\BIBdecl}{\relax}
\BIBdecl

\bibitem{Allgower:Birkhauser:2000}
F.~Allgower and A.~Zheng, \emph{Nonlinear Model Predictive Control}.\hskip 1em
  plus 0.5em minus 0.4em\relax Birkh{\"a}user Basel, 2000.

\bibitem{Qin:CEP:2003}
S.~Qin and T.~A. Badgwell, ``A survey of industrial model predictive control
  technology,'' \emph{Control Engineering Practice}, vol.~11, no.~7, pp. 733 --
  764, 2003.

\bibitem{Rakovic:Birkhauser:2019}
S.~V. Rakovi\'{c} and W.~S. Levine, \emph{Handbook of Model Predictive
  Control}.\hskip 1em plus 0.5em minus 0.4em\relax Birkh{\"a}user Basel, 2019.

\bibitem{Johansen:NTNU:2011}
T.~A. Johansen, ``Introduction to nonlinear model predictive control and moving
  horizon estimation,'' in \emph{Selected Topics on Constrained and Nonlinear
  Control}.\hskip 1em plus 0.5em minus 0.4em\relax STU Bratislava - NTNU
  Trondheim, 2011, ch.~5, pp. 187--233.

\bibitem{Li:IECR:1988}
W.~C. Li and L.~T. Biegler, ``Process control strategies for constrained
  nonlinear systems,'' \emph{Industrial \& Engineering Chemistry Research},
  vol.~27, no.~8, pp. 1421--1433, 1988.

\bibitem{Brengel:IECR:1989}
D.~D. Brengel and W.~D. Seider, ``Multistep nonlinear predictive controller,''
  \emph{Industrial \& Engineering Chemistry Research}, vol.~28, no.~12, pp.
  1812--1822, 1989.

\bibitem{Diehl:Springer:2009}
M.~Diehl, H.~J. Ferreau, and N.~Haverbeke, \emph{Efficient Numerical Methods
  for Nonlinear MPC and Moving Horizon Estimation}.\hskip 1em plus 0.5em minus
  0.4em\relax Springer Berlin Heidelberg, 2009, pp. 391--417.

\bibitem{Rawlings:NHP:2017}
J.~B. Rawlings, D.~Q. Mayne, and M.~Diehl, \emph{Model predictive control :
  theory, computation, and design}, 2nd~ed.\hskip 1em plus 0.5em minus
  0.4em\relax Nob Hill Publishing, 2017.

\bibitem{Ohtsuka:AUTO:2004}
T.~Ohtsuka, ``A continuation/{GMRES} method for fast computation of nonlinear
  receding horizon control,'' \emph{Automatica}, vol.~40, no.~4, pp. 563 --
  574, 2004.

\bibitem{Shen:TCST:2017}
C.~{Shen}, B.~{Buckham}, and Y.~{Shi}, ``Modified {C/GMRES} algorithm for fast
  nonlinear model predictive tracking control of {AUVs},'' \emph{IEEE
  Transactions on Control Systems Technology}, vol.~25, no.~5, pp. 1896--1904,
  2017.

\bibitem{Wang:TCST:2010}
Y.~{Wang} and S.~{Boyd}, ``Fast model predictive control using online
  optimization,'' \emph{IEEE Transactions on Control Systems Technology},
  vol.~18, no.~2, pp. 267--278, 2010.

\bibitem{Biegler:Birkhauser:2000}
L.~T. Biegler, ``Efficient solution of dynamic optimization and nmpc
  problems,'' in \emph{Nonlinear Model Predictive Control}.\hskip 1em plus
  0.5em minus 0.4em\relax Basel: Birkh{\"a}user Basel, 2000, pp. 219--243.

\bibitem{pso}
X.~Wang and J.~Xiao, ``\uppercase{PSO}-based model predictive control for
  nonlinear processes,'' in \emph{Advances in Natural Computation}.\hskip 1em
  plus 0.5em minus 0.4em\relax Springer Berlin Heidelberg, 2005, pp. 196--203.

\bibitem{Du:TCST:2000}
X.~{Du}, K.~K.~K. {Htet}, and K.~K. {Tan}, ``Development of a
  genetic-algorithm-based nonlinear model predictive control scheme on velocity
  and steering of autonomous vehicles,'' \emph{IEEE Transactions on Industrial
  Electronics}, vol.~63, no.~11, pp. 6970--6977, 2016.

\bibitem{seqMC}
A.~Doucet, N.~De~Freitas, and N.~Gordon,
  \emph{\BIBforeignlanguage{English}{Sequential Monte Carlo Methods in
  Practice}}.\hskip 1em plus 0.5em minus 0.4em\relax Springer New York ;
  London, 2001.

\bibitem{sarkka}
S.~Särkkä, \emph{Bayesian Filtering and Smoothing}.\hskip 1em plus 0.5em
  minus 0.4em\relax Cambridge University Press, 2013.

\bibitem{STAHL}
D.~Stahl and J.~Hauth, ``\uppercase{PF-MPC}: Particle filter-model predictive
  control,'' \emph{Systems \& Control Letters}, vol.~60, no.~8, pp. 632--643,
  2011.

\bibitem{Botchu:IFAC:2007}
S.~K. Botchu and S.~Ungarala, ``Nonlinear model predictive control based on
  sequential monte carlo state estimation,'' \emph{IFAC Proceedings Volumes},
  vol.~40, no.~5, pp. 29 -- 34, 2007, 8th IFAC Symposium on Dynamics and
  Control of Process Systems.

\bibitem{Blackmore:AIAA:2006}
L.~Blackmore, \emph{A Probabilistic Particle Control Approach to Optimal,
  Robust Predictive Control}, 2006, pp. AIAA 2006--6240.

\bibitem{Kantas2009}
N.~Kantas, J.~M. Maciejowski, and A.~Lecchini-Visintini, \emph{Sequential Monte
  Carlo for Model Predictive Control}.\hskip 1em plus 0.5em minus 0.4em\relax
  Springer Berlin Heidelberg, 2009, pp. 263--273.

\bibitem{MCMC}
J.~M. Maciejowski, A.~L. Visintini, and J.~Lygeros, \emph{NMPC for Complex
  Stochastic Systems Using a Markov Chain Monte Carlo Approach}.\hskip 1em plus
  0.5em minus 0.4em\relax Springer Berlin Heidelberg, 2007, pp. 269--281.

\bibitem{Rao:TAC:2003}
C.~V. {Rao}, J.~B. {Rawlings}, and D.~Q. {Mayne}, ``Constrained state
  estimation for nonlinear discrete-time systems: stability and moving horizon
  approximations,'' \emph{IEEE Transactions on Automatic Control}, vol.~48,
  no.~2, pp. 246--258, 2003.

\bibitem{Ma-Chen-Flammarion-Jordan:PNAS:2019}
Y.-A. Ma, Y.~Chen, C.~Jin, N.~Flammarion, and M.~I. Jordan, ``Sampling can be
  faster than optimization,'' \emph{Proceedings of the National Academy of
  Sciences}, vol. 116, no.~42, pp. 20\,881--20\,885, 2019.

\bibitem{resample}
G.~Kitagawa, ``Monte carlo filter and smoother for non-gaussian nonlinear state
  space models,'' \emph{Journal of Computational and Graphical Statistics},
  vol.~5, no.~1, pp. 1--25, 1996.

\bibitem{borrelli_bemporad_morari_2017}
F.~Borrelli, A.~Bemporad, and M.~Morari, \emph{Predictive Control for Linear
  and Hybrid Systems}.\hskip 1em plus 0.5em minus 0.4em\relax Cambridge
  University Press, 2017.

\bibitem{lang}
L.~Lang, W.~shiang Chen, B.~R. Bakshi, P.~K. Goel, and S.~Ungarala, ``Bayesian
  estimation via sequential monte carlo sampling—constrained dynamic
  systems,'' \emph{Automatica}, vol.~43, no.~9, pp. 1615--1622, 2007.

\bibitem{copt}
X.~Shao, B.~Huang, and J.~M. Lee, ``Constrained bayesian state estimation – a
  comparative study and a new particle filter based approach,'' \emph{Journal
  of Process Control}, vol.~20, no.~2, pp. 143 -- 157, 2010.

\bibitem{rajmani}
R.~Rajamani, \emph{Vehicle Dynamics and Control}.\hskip 1em plus 0.5em minus
  0.4em\relax Springer US, 2012.

\end{thebibliography}
\end{document}